\documentclass[11pt]{article}
\usepackage{amsmath}
\usepackage{amssymb}
\usepackage{amsthm}
\usepackage{amsfonts}
\usepackage{comment}
\usepackage{color}
\usepackage{mathrsfs}
\usepackage{braket}
\usepackage{graphicx}
\usepackage[subrefformat=parens]{subcaption}
\usepackage{cite}
\usepackage{here}
\usepackage{tikz}
\usetikzlibrary{decorations.pathmorphing}
\usepackage{caption}

\usepackage{listings}
\usepackage{graphicx}

\usepackage[stable]{footmisc}

\makeatletter

\@addtoreset{equation}{section}
\makeatother

\begin{document}

\pagestyle{empty}

\title{\bf{Quasinormal modes of Pleba\'nski--Demia\'nski \\ black hole in the near-Nariai regime}}

\date{}
\maketitle
\begin{center}
{\large 
Hyewon Han\footnote{dwhw101@dgu.ac.kr}, Bogeun Gwak\footnote{rasenis@dgu.ac.kr}
} \\
\vspace*{0.5cm}

{\it 
Department of Physics, Dongguk University, Seoul 04620, Republic of Korea
}

\end{center}

\vspace*{1.0cm}
\begin{abstract}
We investigates the massless scalar perturbations of the Pleba\'nski--Demia\'nski black hole considering the general case that admits all nonzero parameters. This case is the most generic black hole spacetime in general relativity, characterized by mass, spin, acceleration, electric and magnetic charges, NUT parameter, and cosmological constant. Employing conformal transformations, we can separate the massless scalar field equation and reduce the effective potential in the radial perturbation equation into the P\"oschl--Teller potential in the near-Nariai limit where the event and cosmo-acceleration horizons are close. This allows us to obtain an exact analytical solution of the quasinormal frequency, implying that the decay rate of the field is quantized depending only on the surface gravity of the black hole.
\end{abstract}

\newpage
\baselineskip=18pt
\setcounter{page}{2}
\pagestyle{plain}
\baselineskip=18pt
\pagestyle{plain}
\setcounter{footnote}{0}

\section{Introduction}
Black holes have a strong gravitational field and absorb any matter that falls onto them. An event horizon corresponding to the surface of the black hole is a one-way membrane because the matter that has passed through it cannot escape and moves towards the singularity at the center. The structure of a black hole is uniquely determined by its conserved quantities: mass, charge, and angular momentum. A rotating black hole with a nonzero angular momentum forms an ergosphere outside the event horizon. Previous studies have shown that energy can be extracted through the ergosphere \cite{Penrose:1969pc}. However, only part of the energy can be extracted because the black hole has an irreducible mass that does not decrease in any process. The irreducible mass is proportional to the square root of the surface area of the black hole and always increases during the irreversible process \cite{Christodoulou:1970wf,Christodoulou:1971pcn}. This is reminiscent of the second law of classical thermodynamics. Bekenstein defined the entropy of a black hole by relating it to its surface area, describing the black hole system from a thermodynamic perspective \cite{bekenstein1973black}. Furthermore, an analogy between the temperature and the surface gravity of the black hole was suggested. Hawking explained that quantum effects near the event horizon produce pairs of virtual particles, causing the black hole to emit a thermal flux similar to an ordinary body with a temperature proportional to its surface gravity \cite{hawking1975particle}. From these novel discoveries, black hole thermodynamics was established, providing a connection between the classical and quantum properties of black holes.

Studying the propagation and scattering of waves enables us to obtain useful characteristics of the black hole spacetime. This can be achieved at a linearized level by considering the propagating field on a fixed black hole background. In particular, when a gravitational wave packet perturbs a Schwarzschild black hole, the signal is dominated by a damped oscillation mode of a single frequency at intermediate times \cite{Vishveshwara:1970zz}. The oscillation with an exponential decay is the quasinormal mode \cite{Press:1971wr,Chandrasekhar:1975zza,Kokkotas:1999bd,Berti:2009kk,Konoplya:2011qq}, and its associated frequency depends only on the characteristic parameters of the black hole. Furthermore, if the incident wave satisfies certain conditions, then the scattered wave can be amplified. This phenomenon is known as superradiance and causes black hole instability \cite{Press:1972zz,Cardoso:2004nk,Cardoso:2006wa,Brito:2015oca}. The radiation amplification process in a classical dissipative system was discussed in \cite{1971JETPL..14..180Z,1972JETP...35.1085Z}, and superradiant scattering for rotating black holes was considered in \cite{Press:1972zz,Starobinsky:1973aij}. Moreover, assuming confinement, such as the reflecting mirror in the system, the wave has been shown to be exponentially amplified, leading to a black hole bomb. The superradiant instability has been analyzed for astrophysical applications, such as the search for dark matter candidates \cite{Witek:2012tr,Cardoso:2018tly} and hairy black hole solutions \cite{Hod:2012px,Hod:2013zza,Herdeiro:2014goa,Herdeiro:2015gia,Sanchis-Gual:2015lje}.

Research on the quasinormal mode of black holes has developed in various directions, highlighting their physical significance. The main source of current gravitational wave detection is the collision of binary black holes. The ringdown phase, in which the merged black hole settles, is described by the quasinormal mode, and because its frequency is decided solely from the parameters of the final black hole, the identification of the key information of the system is crucial \cite{LIGOScientific:2020tif,Dreyer:2003bv,Capano:2021etf}. Furthermore, the quasinormal frequency contributes to validating the strong cosmic censorship conjecture, which suggests that the singularities of black holes are hidden from all observers and may be related to the determination of the blue-shift instability of the Cauchy horizon \cite{Cardoso:2017soq,Hod:2018lmi,Cardoso:2018nvb,Ge:2018vjq,Gwak:2018rba,Liu:2019lon,Guo:2019tjy,Liu:2019rbq,Mishra:2020gce,Destounis:2020yav,Zhang:2021wda,Casals:2020uxa,Konoplya:2022kld}. Additionally, the anti-de Sitter (AdS)/conformal field theory (CFT) duality \cite{maldacena1999large,MR1633012,aharony2000large} states that the perturbations of a black hole in AdS spacetime, which is characterized by a negative cosmological constant, correspond to the perturbations of a thermal state in the CFT on the AdS boundary. Thus, the timescale for the state in lower-dimensional CFT to reach thermal equilibrium can be predicted by computing the quasinormal modes of the black hole \cite{Horowitz:1999jd,Birmingham:2001pj}.

The quasinormal modes are analyzed by solving perturbations of various test fields, which can be reduced to ordinary differential equations through separation \cite{Teukolsky:1972my}. Because obtaining a solution is typically complicated, various numerical methods have been employed. Interestingly, an analytically exact solution for certain special spacetimes can be obtained. A black hole in de Sitter (dS) spacetime, which is the spacetime of the positive cosmological constant, has a cosmological horizon outside its event horizon. For the near-Nariai black hole where these two horizons are close, the potential in the wave equation is reduced to the P\"oschl--Teller potential \cite{Poschl:1933zz,Ferrari:1984zz}, yielding the general form of the solution. Moreover, the quasinormal frequency analytically obtained by the P\"oschl--Teller approximation agreed well with that computed using numerical methods \cite{Moss:2001ga,Cardoso:2003sw,Suneeta:2003bj,Molina:2003ff,Jing:2003wq,Lopez-Ortega:2006aal}. 

Most objects in the universe rotate. During the coalescence of rotating binary black holes, gravitational waves can be emitted in a specific direction. Then, the final black hole obtains recoil velocity and acceleration \cite{Merritt:2004xa,Bruegmann:2007bri}. The source of the acceleration of the black hole is also explained by the tensions of the cosmic strings \cite{Hawking:1995zn}, the topological defect formed during the phase transitions in the early universe. Therefore, accelerating black hole spacetime may be suitable for describing more general and natural gravitational phenomena. The most generic exact solution of the Einstein--Maxwell equation with a cosmological constant representing the accelerating black hole is the Pleba\'nski--Demia\'nski solution. The original form of this metric was proposed in \cite{Plebanski:1976gy} and further improved in \cite{Hong:2004dm,Griffiths:2005qp,Podolsky:2021zwr,Podolsky:2022xxd} with physical interpretations of the parameters and a discussion of various features. The Pleba\'nski--Demia\'nski metric is characterized by mass, spin, acceleration, electric and magnetic charges, NUT parameter, and cosmological constant. This metric encompasses well-known solutions, such as the Kerr--Newman metric, C-metric, and Taub--NUT metric, as its special subcases. For a deeper understanding of the physics in strong gravitational fields, the properties of this extended black hole spacetime must be scrutinized in various aspects. Consequently, the perturbations and quasinormal modes of the Pleba\'nski--Demia\'nski family have been analyzed in recent years \cite{Destounis:2020pjk,Wei:2021bqq,Destounis:2022rpk,Gwak:2022nsi,Xiong:2023usm,BarraganAmado:2023wxt,Chen:2024feh,Chen:2024rov}.

In this study, we investigate the quasinormal modes of an accelerating and spinning charged NUT black hole in asymptotically dS spacetime. We employ the Pleba\'nski--Demia\'nski metric where all parameters are nonzero. The analysis of perturbations on this general black hole remains open. We consider a massless scalar field conformally coupled to the gravitational field and use the conformal transformations, which enable us to easily separate the field equation. The presence of both acceleration and positive cosmological constant results in a cosmo-acceleration horizon outside the event horizon. We focus on the near-Nariai case for which the event horizon and the cosmo-acceleration horizon are closely located. Moreover, we demonstrate that in such spacetime the radial perturbation equation can be reduced to an effective equation with the P\"oschl--Teller potential. Subsequently, the solution of the quasinormal frequency is analytically obtained. We find that the decay rate of the scalar perturbation on the near-Nariai Pleba\'nski--Demia\'nski black hole is quantized for the overtone index $n$, depending on the surface gravity.

The remainder of this paper is organized as follows. In Sec\,2, we briefly review the Pleba\'nski–Demia\'nski black hole. In Sec\,3, we consider a massless scalar perturbation with conformal coupling. The boundary conditions defining the quasinormal mode are presented. In Sec\,4, we analytically derive the quasinormal frequencies for the near-Nariai black hole. Finally, Sec\,5 summarizes the study. In this study, geometrized units, where $c=G=1$, were employed.

\section{Pleba\'nski–Demia\'nski solution}
To consider a general black hole geometry, we use the Pleba\'nski–Demia\'nski solution \cite{Plebanski:1976gy}, which describes a pair of uniformly accelerating and rotating charged black holes with a NUT parameter and a cosmological constant. This section reviews the basic properties of the generic Pleba\'nski–Demia\'nski black hole. We employ an explicit form of the metric presented in \cite{Podolsky:2022xxd} and consider the positive cosmological constant $\Lambda>0$. (Further details of the solution are discussed in \cite{Podolsky:2022xxd} and references therein.) The line elements are given by
    \begin{align} \label{metric1}
        ds^2=\frac{1}{\Omega^2} &\left(- \frac{Q}{\rho^2} \left[dt-\left(a \sin^2\theta +4l\sin^2 \frac{\theta}{2} \right)d\varphi \right]^2 +\frac{\rho^2}{Q} dr^2 +\frac{\rho^2}{P} d\theta^2 \right. \nonumber \\ 
        & \quad \left. +\frac{P}{\rho^2} \sin^2 \theta \left[a dt-\left(r^2+(a+l)^2\right) d\varphi \right]^2\right),
    \end{align}
where
    \begin{align} 
        \Omega&=1-\frac{\alpha a}{a^2+l^2} (l+a\cos\theta)r, \label{conformalfactor} \\ 
        \rho^2&=r^2+(l+a \cos \theta)^2, \\
        P&= 1-2\left( \frac{\alpha a}{a^2+l^2} M -\frac{l}{L^2}\right)(l+a\cos\theta) \nonumber\\
        & \qquad +\left(\frac{\alpha^2 a^2}{(a^2+l^2)^2} (a^2-l^2+e^2+g^2)+\frac{1}{L^2} \right)(l+a\cos\theta)^2, \label{metf2}  \\
        Q&=\left[r^2-2Mr+(a^2-l^2+e^2+g^2)\right]\left(1+\alpha a \frac{a-l}{a^2+l^2} r \right) \left(1-\alpha a \frac{a+l}{a^2+l^2} r \right) \nonumber\\
        & \qquad -\frac{r^2}{L^2}\left[r^2+2\alpha a l \frac{a^2-l^2}{a^2+l^2}r +(a^2+3l^2) \right]. \label{metf}        
    \end{align}
The physical parameters in the metric are as follows: $M$ is the mass, $a$ is the rotation, $\alpha$ is the acceleration, $l$ is the NUT parameter, $e$ and $g$ are the electric and magnetic charges, respectively, and $L=(3/\Lambda)^{1/2}$ is a curvature radius of dS spacetime. The solution portrays the well-known black hole spacetimes of general relativity by properly setting the parameters to zero. For example, when $L^{-2}=l=0$, the solution describes the spinning charged C-metric, and when $\alpha=e=g=0$, the Kerr metric is obtained. In this study, we considered spacetime in which all parameters are nonzero.

An additional parameter is hidden within the angular coordinate range, $\varphi$. The metric \eqref{metric1} admits the deficit or excess angles, as the circumference over the radius around the boundaries of the range of $\theta \in [0, \pi]$ deviates from $2\pi$. They correspond to the conical singularities at the axes of symmetry $\theta=0, \pi$ and uniformly accelerate the black hole by the tension of the cosmic string or the stress of the cosmic strut. Conical singularities can be eliminated by assuming the range of angular coordinate as $\varphi \in [0, 2\pi \mathcal{C})$ and properly selecting the conicity $\mathcal{C}$. When the value of this parameter is specified using the metric function \eqref{metf2} as $\mathcal{C}=1/P(0)$, which corresponds to the selection of regularizing the axis $\theta=0$, the deficit/excess angle on the axis $\theta=\pi$ is given by
        \begin{align} \label{da1}
            \delta_{\pi}&=\frac{-8\pi a [\alpha a \{M(a^2+l^2)-\alpha a l (a^2-l^2+e^2+g^2)\}-\frac{2l}{L^2}(a^2+l^2)^2]}{\{1+\frac{(a+l)(a+3l)}{L^2}\}(a^2+l^2)^2-2\alpha aM(a+l)(a^2+l^2)+\alpha^2a^2(a+l)^2(a^2-l^2+e^2+g^2)}.
        \end{align}
Similarly, when we assume that $\mathcal{C}=1/P(\pi)$, the singularity on the axis $\theta=\pi$ can be removed, and the deficit/excess angle on the axis $\theta=0$ is
        \begin{align} \label{da2}
\delta_{0}&=\frac{8\pi a [\alpha a \{M(a^2+l^2)-\alpha a l (a^2-l^2+e^2+g^2)\}-\frac{2l}{L^2}(a^2+l^2)^2]}{\{1+\frac{(a-l)(a-3l)}{L^2}\}(a^2+l^2)^2+2\alpha aM(a-l)(a^2+l^2)+\alpha^2a^2(a-l)^2(a^2-l^2+e^2+g^2)}.
        \end{align}
If we select particular values of the parameters that satisfy the relation
        \begin{align} \label{conremove}
            \frac{2l}{L^2}(a^2+l^2)^2=\alpha a [M(a^2+l^2)-\alpha a l (a^2-l^2+e^2+g^2)],
        \end{align}
the values of the metric function $P(\theta)$ on both axes are identical, and the deficit/excess angles simultaneously vanish. Therefore, conical singularities disappear, and the black hole accelerates without any physical sources, such as cosmic strings or struts \cite{Podolsky:2022xxd}. Additionally, for rotating and accelerating black holes with the vanishing NUT parameter $l=0$, the conical singularities on both axes cannot be removed simultaneously. Because we consider a general solution in which all parameters are nonzero, we can eliminate both singularities that occur at the axes $\theta=0$ and $\theta=\pi$ by requiring that the constraint \eqref{conremove} is satisfied and $\mathcal{C}=1/P(0)=1/P(\pi)$.

The spacetime represented by metric \eqref{metric1} exhibits a curvature singularity at $\rho^2=0$, which requires both $r=0$ and $l+a \cos \theta=0$. We consider a black hole containing the curvature singularity by assuming that the values of the rotation $a$ and NUT $l$ parameters satisfy the relation $\left| a \right| \ge \left| l \right|$. The horizons of the black hole are located at $r_h$, such that $Q(r_h)=0$. This quartic equation can have maximally four distinct real roots under the condition
    \begin{align}
        \frac{1}{L^2}\ne -\alpha^2 a^2 \frac{a^2-l^2}{(a^2+l^2)^2},
    \end{align}
which corresponds to the general case. We rewrite $Q(r)$ in a factorized form as follows
    \begin{align} \label{fm}
        Q(r)=-\mathcal{N} (r-r_b^+)(r-r_b^-)(r-r_c^+)(r-r_c^-),
    \end{align}
where
    \begin{align}
        \mathcal{N}=\alpha^2 a^2 \frac{a^2-l^2}{(a^2+l^2)^2}+\frac{1}{L^2}.
    \end{align}
The roots $r_b^+, r_b^-, r_c^+$, and $r_c^-$ are the locations of the outer black hole, inner black hole, outer cosmo-acceleration, and inner cosmo-acceleration horizons, respectively. We employ a natural ordering for the horizons as $r_c^- < r_b^- < r_b^+ < r_c^+$ by assuming $\mathcal{N}>0$ (note that this is automatically satisfied if the condition for singularity $\left| a \right| \ge \left| l \right|$ is satisfied). In this case, $r=r_b^+$ corresponds to the event horizon, and $r=r_c^-$ lies in the negative $r$ region, which is not considered. Thus, we call $r_c^+$ the position of the cosmo-acceleration horizon for simplicity. This study focuses on the region between the event and cosmo-acceleration horizons $r_b^+<r<r_c^+$, which may be a stationary and observable region outside the black hole. 

Furthermore, the thermodynamic variables of the black hole can be defined on each horizon. The temperature and the entropy are given by
        \begin{align}
             T= \frac{\hbar}{2 \pi k_B} \kappa, \qquad S =  \frac{k_B}{4 \hbar} A,
        \end{align}
where $\hbar$ and $k_B$ denote the reduced Planck and Boltzmann constants, respectively. They are determined by the surface gravity $\kappa$ and the surface area $A$, which are calculated on the corresponding horizons $r_h$ as follows
        \begin{align}
            \kappa = \frac{\left. \partial_r Q \right|_{r=r_h} }{2(r_h^2+(a+l)^2)}, \qquad A =  \frac{4 \pi \mathcal{C}(r_h^2+(a+l)^2) }{(1-\alpha a \frac{a+l}{a^2+l^2}r_h)(1+\alpha a \frac{a-l}{a^2+l^2}r_h)}.
        \end{align}
The point where the horizon has vanishing surface gravity corresponds to an extremal horizon, implying that two or more horizons coincide. A black hole with three horizons in the positive $r$ region, which entails three extreme cases, is considered. We analyze the quasinormal modes of the near-extreme black hole in the following sections.

\section{Massless scalar field perturbation}
We examine the perturbations of a neutral massless scalar field. The field equation on the Pleba\'nski–Demia\'nski metric \eqref{metric1} is slightly complicated due to the existence of the conformal factor $\Omega(r,\theta)$. Nevertheless, the equation can be separated through conformal transformations \cite{Destounis:2020pjk,Gwak:2022nsi,BarraganAmado:2023wxt}. We demonstrate that this can be also achieved in the general Pleba\'nski–Demia\'nski spacetime. By introducing a massless scalar field $\Psi$, which is conformally coupled with the gravitational field, we have the conformally invariant Klein--Gordon equation \cite{Wald:1984rg}
    \begin{align} \label{we1}
        g^{\mu \nu}\nabla_\mu \nabla_\nu \Psi -\frac{1}{6} R \Psi =0,
    \end{align}
where $R=12/L^2$ is the Ricci scalar of the metric \eqref{metric1}. To separate the field equation, we considered a simpler form of the metric in which the conformal factor is eliminated. By performing the conformal transformations 
    \begin{align}
        \tilde{g}_{\mu \nu} = \Omega^2 g_{\mu \nu}, \qquad \tilde{\Psi} = \Omega^{-1} \Psi,
    \end{align}
the metric \eqref{metric1} is scaled to
    \begin{align}
        d\tilde{s}^2=- &\frac{Q}{\rho^2} \left[dt-\left(a \sin^2\theta +4l\sin^2 \frac{\theta}{2} \right)d\varphi \right]^2 +\frac{\rho^2}{Q} dr^2 +\frac{\rho^2}{P} d\theta^2  \nonumber \\ 
        & \quad  +\frac{P}{\rho^2} \sin^2 \theta \left[a d t-\left(r^2+(a+l)^2\right) d\varphi \right]^2.
    \end{align}
The field equation \eqref{we1} also transforms to
    \begin{align} \label{confe}
        \frac{1}{\sqrt{-\tilde{g}}} \partial_\mu \left( \sqrt{-\tilde{g}} \tilde{g}^{\mu \nu} \partial_{\nu} \tilde{\Psi} \right) -\frac{1}{6} \tilde{R} \tilde{\Psi} =0,
    \end{align}
where 
    \begin{align}
        \sqrt{-\tilde{g}}= \rho^2 \sin\theta, \qquad \tilde{R}=-\frac{1}{\rho^2}\left( -2P+3\dot P \cot\theta +\ddot P+Q'' \right).
    \end{align}
The dot and prime symbols denote the differentiation with respect to $\theta$ and $r$, respectively. We consider the ansatz for the conformally scaled scalar field
    \begin{align}
        \tilde{\Psi}(t,r,\theta,\varphi)=e^{-i\omega t}e^{im\varphi}\psi(r) \Theta(\theta),
    \end{align}
where $\omega$ is the quasinormal frequency, and $m$ is the azimuthal quantum number. Because we have redefined the range of the periodic coordinate $\varphi$ by selecting the parameter $\mathcal{C}=1/P(0)=1/P(\pi)$ under \eqref{conremove}, $m$ must have a form of $m=m_0 P(0)=m_0 P(\pi)$ with integer $m_0$ \cite{Destounis:2020pjk}. Using the ansatz, Eq.\eqref{confe} can be separated as follows
    \begin{align}
        &\frac{1}{\psi}\partial_r[Q\partial_r \psi]+ \frac{\left[\omega (r^2+(a+l)^2)-ma\right]^2}{Q} +\frac{Q''}{6}-\mathcal{K}=0, \label{radial} \\
        &\frac{1}{\Theta \sin\theta} \partial_\theta[P \sin\theta \partial_\theta \Theta]- \frac{\left[\omega(a\sin^2\theta+4l\sin^2\frac{\theta}{2})-m\right]^2}{P \sin^2\theta}+\frac{-2P+3\dot P \cot\theta+\ddot P}{6}+\mathcal{K} =0, \label{ang}
    \end{align}
where $\mathcal{K}$ is a separation constant. We now employ appropriate boundary conditions defining the quasinormal mode. By introducing
    \begin{align} \label{tor}
        \frac{d r_*}{dr}=\frac{r^2+(a+l)^2}{Q}, \qquad \psi=\frac{\tilde{\psi}}{\sqrt{r^2+(a+l)^2}},
    \end{align}
the separated radial equation \eqref{radial} is rewritten as follows
    \begin{align} \label{radial2}
        &\frac{d^2\tilde{\psi}(r_*)}{dr_*^2}+\left[\left(\omega-\frac{m a}{r^2+(a+l)^2} \right)^2 +\frac{Q Q''}{6(r^2+(a+l)^2)^2}-\frac{Q \mathcal{K}}{(r^2+(a+l)^2)^2} \right. \nonumber \\
        & \qquad \qquad \qquad \left. -\frac{rQ Q'}{(r^2+(a+l)^2)^3}-\frac{Q^2((a+l)^2-2r^2)}{(r^2+(a+l)^2)^4} \right]\tilde{\psi}(r_*)=0.
    \end{align}
Because we restrict our analysis to the region $r_b^+<r<r_c^+$, the boundaries of the radial coordinate $r$ correspond to horizons $r=r_b^+$ and $r=r_c^+$, such that $Q(r_b^+)=Q(r_c^+)=0$. At the horizons, Eq.\eqref{radial2} becomes
    \begin{align}
        &\frac{d^2\tilde{\psi}(r_*)}{dr_*^2}+(\omega-m\Omega_b)^2 \tilde{\psi}(r_*) =0, \qquad r\to r_b^+, \\
        &\frac{d^2\tilde{\psi}(r_*)}{dr_*^2}+(\omega-m\Omega_c)^2 \tilde{\psi}(r_*) =0, \qquad r\to r_c^+,
    \end{align}
where $\Omega_b$ and $\Omega_c$ are the angular velocities of the event and cosmo-acceleration horizons, respectively,
    \begin{align}
        \Omega_b=\frac{a}{r_b^{+2} +(a+l)^2}, \qquad \Omega_c=\frac{a}{r_c^{+2} +(a+l)^2}.
    \end{align}
Subsequently, the solutions at the horizons are obtained by
    \begin{align}
        \tilde{\psi}(r\to r_b^+)& \sim e^{- i (\omega-m\Omega_b)r_*}, \qquad \mathrm{(ingoing)} \label{asym1} \\
        \tilde{\psi}(r\to r_c^+)& \sim e^{+ i (\omega-m\Omega_c)r_*}. \qquad \mathrm{(outgoing)} \label{asym2}
    \end{align}
These conditions ensure that only ingoing modes are allowed in the event horizon, whereas only outgoing modes exist in the cosmo-acceleration horizon. For the angular part, we use the following transformation
    \begin{align}
        dz=\frac{d\theta}{P\sin\theta}.
    \end{align}
Then, the separated angular Eq.\eqref{ang} can be expressed by
    \begin{align} \label{ang2}
        \frac{d^2 \Theta(z)}{dz^2} - & \left[ m^2 +\omega^2 \left(a\sin^2\theta+4l\sin^2\frac{\theta}{2}\right)^2-2\omega m \left(a\sin^2\theta+4l\sin^2\frac{\theta}{2}\right) \right. \nonumber \\ 
        & \quad \left.-\frac{P\sin^2\theta}{6}(-2P+3\dot P \cot\theta+\ddot P) -\mathcal{K}P\sin^2\theta \right] \Theta(z) =0,
    \end{align}
and the boundaries $\theta=0$ and $\theta=\pi$ correspond to $z=\mp \infty$, respectively. At the boundaries, we obtain
    \begin{align}
        \frac{d^2 \Theta(z)}{dz^2} - m^2 \Theta(z) =0, \qquad &z \to -\infty, \\
         \frac{d^2 \Theta(z)}{dz^2} - (m-4 l \omega )^2 \Theta(z)=0, \qquad &z \to +\infty.
    \end{align}
We require a regular solution on the boundaries of the range of $\theta$; thus, taking the boundary conditions of the angular function as follows
    \begin{align} 
        \Theta(z) \sim e^{+ \left| m \right| z}, \qquad &z=-\infty \, (\theta = 0), \label{asym3} \\
         \Theta(z) \sim e^{- \left| m-4l \omega  \right|z}, \qquad &z=+\infty \, (\theta = \pi). \label{asym4}
    \end{align}
The difference in the exponents between the solutions at each boundary is because the nonvanishing NUT parameter produces a difference in the magnitude of rotation of each axis. When evaluating the angular velocity parameter $\omega_\theta \equiv g_{t\varphi}/g_{tt}$ near both axes, the axis $\theta=0$ does not rotate, i.e., $\omega_0=0$, and the axis $\theta=\pi$ rotates with angular velocity $\omega_\pi=-4l$. 

The asymptotic solutions \eqref{asym1}--\eqref{asym2} and \eqref{asym3}--\eqref{asym4} are the physically motivated boundary conditions for the quasinormal modes. The oscillatory frequencies and decay rates of the scalar fields on the Pleba\'nski–Demia\'nski black holes can be computed by solving the perturbation equations \eqref{radial2} and \eqref{ang2} with the conditions.

\section{Quasinormal modes in the near-Nariai limit}
Various approximation schemes or numerical methods have been used to compute the quasinormal modes of black holes. The P\"oschl--Teller approximation \cite{Poschl:1933zz,Ferrari:1984zz} permits analytical calculations to obtain exact solutions near the Nariai regime. In this section, we analytically derive the quasinormal frequency for the near-Nariai Pleba\'nski–Demia\'nski black hole.

\begin{figure}[h!] 
\noindent\begin{subfigure}[b]{0.5\textwidth}
    \centering
    \includegraphics[scale=0.8]{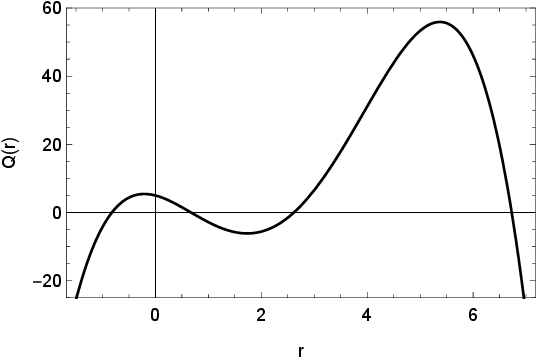}
    \caption{Nonextreme black hole}
\end{subfigure}%
\noindent\begin{subfigure}[b]{0.5\textwidth}
    \centering
    \includegraphics[scale=0.8]{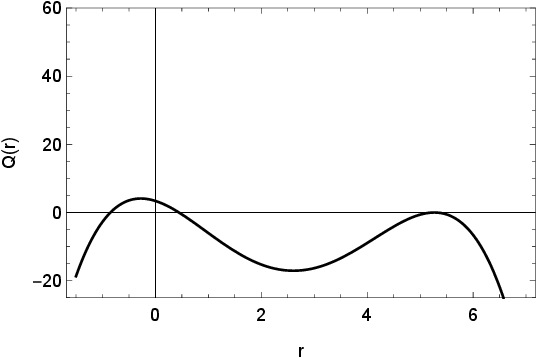}
    \caption{Nariai black hole}
\end{subfigure} \\

\noindent\begin{subfigure}[b]{0.5\textwidth}
    \centering
    \includegraphics[scale=0.8]{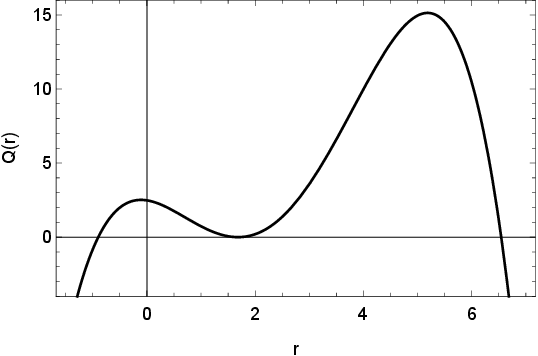}
    \caption{Kerr-type extreme (cold) black hole}
\end{subfigure}%
\noindent\begin{subfigure}[b]{0.5\textwidth}
    \centering
    \includegraphics[scale=0.8]{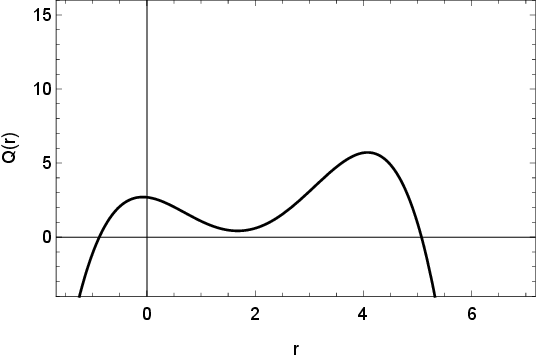}
    \caption{Naked singularity}
\end{subfigure}%
\caption{Metric function $Q$ of the Pleba\'nski–Demia\'nski metric with respect to the radial coordinate $r$ for $\alpha=e=g=l=1$ and $\Lambda=0.1$. The values $r$ where $Q=0$ correspond to the horizon positions. (a) For $M=4$ and $a=-2$, the black hole has three distinct horizons in the region of positive $r$. (b) Nariai black hole with $M=4$ and $a=-1.54645 \cdots$. (c) Kerr-type extreme black hole (cold black hole) with $M=1.5$ and $a=-1.21716 \cdots$. (d) For $M=1.5$ and $a=-1.3$, the singularity is naked.} \label{fig}
\end{figure}

The generic Pleba\'nski–Demia\'nski black hole has three distinct horizons in the positive $r$ region. In this spacetime, two extreme cases exist in which the sizes of the two horizons are identical: the Kerr-type extreme black hole (or cold black hole) where the event horizon $r_b^+$ coincides with the inner black hole horizon $r_b^-$, and the Nariai black hole where the cosmo-acceleration horizon $r_c^+$ coincides with the event horizon $r_b^+$ (additionally, the ultracold black hole exists where all three horizons coincide). The metric functions $Q(r)$ \eqref{metf} corresponding to each extreme case are shown in Figure \ref{fig}. We considered the black hole spacetime satisfying conditions close to those of the Nariai case, as shown in Figure \ref{fig}(b).

The event and cosmo-acceleration horizons of the near-Nariai Pleba\'nski–Demia\'nski black hole are located extremely close. That is, $r_b^+ \approx r_c^+$. Then, the stationary region outside the black hole is confined to a small range of regions $r_b^+ \lesssim r \lesssim r_c^+$ and is characterized by $Q(r) \ll 1$ and $Q'(r) \ll 1$. If we only consider the first orders of $Q$ and $Q'$, the radial equation, Eq.\eqref{radial2}, of the near-Nariai black hole is obtained by
    \begin{align} \label{ner}
        \frac{d^2\tilde{\psi}(r_*)}{dr_*^2}+\left[\left(\omega-\frac{m a}{r^2+(a+l)^2} \right)^2 +\frac{Q Q''}{6(r^2+(a+l)^2)^2}-\frac{Q \mathcal{K}}{(r^2+(a+l)^2)^2} \right]\tilde{\psi}(r_*)=0.
    \end{align}
Using the factorized form Eq.\eqref{fm} of the metric function, we have
    \begin{align} 
        \left. Q'(r) \right|_{r=r_b^+} \equiv Q'_b=\mathcal{N} (r_c^+-r_b^+)(r_b^+-r_b^-)(r_b^+-r_c^-),
    \end{align}
yielding the surface gravity on the event horizon as
    \begin{align} 
        \kappa(r_b^+) \equiv \kappa_b =\frac{Q'_b}{2(r_b^{+2}+(a+l)^2)}=\frac{\mathcal{N} (r_c^+-r_b^+)(r_b^+-r_b^-)(r_b^+-r_c^-)}{2(r_b^{+2}+(a+l)^2)}.
    \end{align}
By considering the tortoise coordinate in the vicinity of the event horizon,
    \begin{align} 
        r_*=\int \frac{r_b^{+2}+(a+l)^2}{Q} dr=\frac{1}{2\kappa_b} \ln \left( \frac{r-r_b^+}{r_c^+-r}\right),
    \end{align}
the radial coordinate such that $r_b^+ \lesssim r \lesssim r_c^+$ is expressed as follows
    \begin{align} 
        r=\frac{r_b^+ +r_c^+ e^{2\kappa_b r_*}}{1+e^{2\kappa_b r_*}}.
    \end{align}
Then, the metric function \eqref{fm} can be expressed in terms of the surface gravity and tortoise coordinate as
    \begin{align} 
        Q=\frac{\kappa_b (r_c^+ -r_b^+)(r_b^{+2}+(a+l)^2) }{2 \cosh^2(\kappa_b r_*) }.
    \end{align}
Subsequently, Eq.\eqref{ner} takes the form
    \begin{align} 
        \frac{d^2\tilde{\psi}}{dr_*^2}+\left[\left(\omega-m\Omega_b \right)^2 -\frac{V_0}{\cosh^2(\kappa_b r_*)} \right]\tilde{\psi}=0,
    \end{align}
under the near-Nariai conditions. This corresponds to a wave equation with an effective potential of the P\"oschl--Teller type 
    \begin{align} 
        V_0=\frac{\kappa_b (r_c^+ -r_b^+)}{2(r_b^{+2}+(a+l)^2)}\left(\mathcal{K}-\frac{Q''_b}{6} \right), 
    \end{align}
where
    \begin{align} 
        Q''_b \equiv & \left. Q''(r) \right|_{r=r_b^+} = -12 \left[\alpha^2 a^2 \frac{a^2-l^2}{(a^2+l^2)^2}+\frac{1}{L^2} \right] r_b^{+2} \nonumber \\
        & \quad +12 \alpha a \left[\alpha a M \frac{a^2-l^2}{(a^2+l^2)^2}-\frac{l}{a^2+l^2}-\frac{a^2-l^2}{a^2+l^2}\frac{l}{L^2} \right] r_b^+ \nonumber \\
        & \quad +2 \left[1+4\alpha a M \frac{l}{a^2+l^2} - \alpha^2 a^2 \frac{a^2-l^2}{(a^2+l^2)^2} (a^2-l^2+e^2+g^2)-(a^2+3l^2) \frac{1}{L^2}\right].
    \end{align}
The exact solution for the P\"oschl--Teller potential was obtained in \cite{Ferrari:1984zz,Cardoso:2003sw} as follows
    \begin{align} 
        \mathrm{Re}[\omega]&=m\Omega_b + \kappa_b \sqrt{\frac{1}{2 \kappa_b} \left(\frac{r_c^+ -r_b^+}{r_b^{+2}+(a+l)^2}\right) \left(\mathcal{K}- \frac{Q''_b}{6} \right) -\frac{1}{4}}, \label{real} \\
        \mathrm{Im}[\omega]&=-\left(n+\frac{1}{2} \right) \kappa_b, \qquad n=0,1,2,\cdots, \label{imaginary}
    \end{align}
where $n$ is the overtone number. We show that the quasinormal frequency of the near-Nariai Pleba\'nski–Demia\'nski black hole for massless scalar field with the conformal coupling can be analytically found by the P\"oschl--Teller approximation. The real part of Eq.\eqref{real} of the complex frequency is associated with field oscillation. In contrast, the imaginary part of Eq.\eqref{imaginary} determines the damping time, and its magnitude depends on the surface gravity of the black hole and the non-negative integer $n$ that labels each quantized mode. The $n=0$ mode corresponds to the fundamental mode with the least damping, which typically dominates the quasinormal mode. This form of Eq.\eqref{imaginary} has been obtained in previous studies \cite{Cardoso:2003sw,Molina:2003ff,Gwak:2019ttv,Gwak:2022nsi} as a result of perturbations in the near-extreme black holes. Notably, even for the general black hole spacetime with nonzero acceleration and the NUT parameter, the decay rate of the field is given in such a universal form in the near-Nariai regime.

\section{Conclusions}
The quasinormal modes of scalar perturbations on near-Nariai Pleba\'nski–Demia\'nski black hole were analytically studied. We considered an accelerating and rotating charged NUT black hole in asymptotically dS spacetime, which was described by the general Pleba\'nski–Demia\'nski metric. We examined a massless scalar field conformally coupled to gravity and performed transformations to remove the conformal factor in the metric, which disturbs the separation of the field equation. The conformally invariant Klein--Gordon equation was separated into radial and angular equations using the scaled metric, and physically related quasinormal boundary conditions were specified.

The Pleba\'nski–Demia\'nski black hole has a cosmo-acceleration horizon outside its event horizon due to both acceleration and positive cosmological constant. We focused on the near-Nariai condition where these two horizons are located close. In such spacetime, the Schrödinger-type equation with the P\"oschl--Teller potential could be obtained. The equation provides an analytical solution of the quasinormal frequency. Consequently, we observed that the decay rate of the field depended only on the surface gravity of the black hole for each mode $n$.

\vspace{10pt} 

\noindent{\bf Acknowledgments}

\noindent This research was supported by Basic Science Research Program through the National Research Foundation of Korea (NRF) funded by the Ministry of Education (NRF-2022R1I1A2063176) and the Dongguk University Research Fund of 2024. BG appreciates APCTP for its hospitality during completion of this work.\\

\bibliographystyle{bibstyle}
\bibliography{ref}
\end{document}